# Transit time of a freely-falling quantum particle in a background gravitational field

P.C.W. Davies

Australian Centre for Astrobiology, Macquarie University, New South Wales 2109, Australia

**Abstract**

Using a model quantum clock, I evaluate an expression for the time of a non-relativistic quantum particle to transit a piecewise geodesic path in a background gravitational field with small spacetime curvature (gravity gradient), in the case that the apparatus is in free fall. This calculation complements and extends an earlier one (Davies 2004) in which the apparatus is fixed to the surface of the Earth. The result confirms that, for particle velocities not too low, the quantum and classical transit times coincide, in conformity with the principle of equivalence. I also calculate the quantum corrections to the transit time when the de Broglie wavelengths are long enough to probe the spacetime curvature. The results are compared with the calculation of Chaio and Speliotopoulos (2003), who propose an experiment to measure the foregoing effects.

**1. Background**

The quantum mechanics of particles moving in a background gravitational field has become of considerable experimental and theoretical interest (Collela *et al* 1975, Mashhoon 1988, Amino *et al* 1993, Peters *et al* 1999, van der Zouw *et al* 2000). In most analyses, the particles are treated as sharply-peaked non-relativistic wave packets, in which case their motion approximates the classical limit. But one may also ask about the behaviour of highly non-classical quantum states, raising important issues about the consistency of quantum mechanics and the theory of relativity, two subjects that proceed from very different conceptual bases. In particular, the theory of relativity is founded on the use of measuring rods and clocks, an essentially classical concept that is challenged by quantum uncertainty, and other non-classical properties such as entanglement.

One approach for investigating the intersection of the theory of relativity with highly non-classical quantum states is to use a model quantum clock to measure the time of flight of quantum particles. Such a clock was introduced many years ago by Salecker and Wigner (1958), and elaborated by Peres (1980). It involves computing $\varphi$, the change in the phase of the wave function, over the duration to be measured. The transit time is then given by

$$T = \hbar \partial \varphi / \partial E \qquad (1.1)$$

where $E$ is the particle's energy. This model has the virtue that it yields precise expectation values for the transit time of quantum particles passing through a prescribed region of space, even though the absolute time of passage suffers from quantum uncertainty (Peres 1980). This remains true for highly spatially delocalized states such as energy eigenstates. For example, a uniform flux of particles of mass $m$ in free space, described by plane wave energy and momentum eigenstates, $e^{ikx-iEt/\hbar}$, predicts a transit time through the region $0 \leq x \leq d$ of $d/v$, where $v \equiv \hbar k/m$, in analogy to the classical velocity.

In an earlier paper (Davies 2004), I applied the quantum clock analysis to the motion of a quantum particle in a uniform gravitational field – a scenario that has been the subject of an extensive experimental programme (see, for example, Amino *et al* 1993). The results showed that the transit time of a particle projected vertically and allowed to fall back again approaches the classical transit time when the measurement takes place far from the turning point using a clock fixed in the reference frame of the Earth. Thus the quantum particle obeys the principle of equivalence (in this approximate sense), in spite of the fact that there is a mass-dependant probability that the particle may suffer 'quantum delay' by tunnelling into the classically forbidden region above the classical turning point. It turns out that the quantum delay is exactly cancelled by the probability that the particle may back-scatter off the gravitational potential $V(x) = mgx$ before reaching the classical turning point.

In this paper I consider the case of a non-uniform gravitational field. The previous treatment may readily be adapted to include a small gravity gradient term in the potential, but rather than present that analysis here, I shall treat instead the case in which the quantum system, including the clock, is in free fall. This enables one to eliminate the linear $mgx$ term in the potential, and to work with simple plane wave eigenstates rather than the more complicated Airy functions appropriate for the uniform gravitational potential. The non-trivial transformation between the two sets of eigenstates under a uniform acceleration has been considered by Berry and Balazs (1979).

**2. Analysis in the freely falling frame**

The gravitational potential of the Earth at a distance $x$ above the surface may be approximated as

$$- GMm/(R + x) \approx - GMm/R + GMmx/R^2 - GMmx^2/R^3 \tag{2.1}$$

where $M$ and $R$ are the mass and radius of the Earth respectively. The first term on the right hand side is a constant potential and may be scaled out. The second, linear, term is the uniform $mgx$ potential, which we shall transform away by working in a freely-falling frame. The third term, quadratic in $x$, will provide the potential in which the quantum particle moves. The Schrödinger equation to be solved is thus

$$- (\hbar^2/2m)\partial^2\psi/\partial x^2 + \alpha(x+b)^2\psi = i\hbar\partial\psi/\partial t \tag{2.2}$$

where

$$\alpha \equiv -GMm/R^3 \tag{2.3}$$

and for later convenience I have shifted the origin of the spatial coordinate: $x \to x + b$, ($b > 0$).

The situation I shall consider is shown in Fig. 1. The freely-falling frame contains the clock, which is at rest relative to the 'start line' at $x = -b$. The particle propagates in a direction away from Earth with energy $E$ relative to the frame of the clock, triggering the clock mechanism when it crosses the start line. The tidal gravitational force will serve to increase the separation of the clock and the particle, so for ready comparison with the 'out-and-back' transit time analysis of the earlier paper (Davies 2004), I introduce a reflecting barrier at $x = 0$, at rest relative to the clock, to reflect the particle back again. The clock switches off when the particle returns to the start line. The clock thus measures the time for the particle to propagate to the barrier and back again, in the freely-falling frame. The artifice of the reflecting wall enables me to simplify the analysis by treating motion in one space dimension: it may be dispensed with by considering orbital motion in a plane (see the discussion in section 3).

The potential corresponding to this set-up is

$$V(x) = 0 \qquad x < -b$$
$$= \alpha(x+b)^2 \qquad -b \leq x < 0 \tag{2.4}$$
$$= \infty \qquad x = 0$$

and is shown in Fig. 2. Note that I am free to put $V(x) = 0$ for $x < -b$, because, by assumption, the clock will not run when the particle is in that region; we may restrict the analysis to $x > -b$.

In practice, the tidal forces would be a very small perturbation on the free particle motion, enabling the use of perturbation theory. The classical out-and-back transit time is given by

$$T_{\text{classical}} = (2m)^{1/2} \int_{-b}^{0} dx/(E-V)^{1/2} \approx 2b/v + (1/Ev)\int_{-b}^{0} V(x')dx'. \tag{2.5}$$

for $V \ll E$. The first term on the right hand side of Eq. (2.5) is the transit time for a free particle to go from $x = -b$ to 0 and back with uniform velocity $v$; the second term is the (small) correction due to the gravity gradient. Denoting the former by $T_0$, and putting $V(x) = \alpha(x+b)^2$, we find for the classical transit time

$$T_{\text{classical}} = T_0 + T_0 b^2 \alpha/6E = T_0 - T_0 b^2 GMm/6R^3 E. \tag{2.6}$$

Note that the final (gravity gradient correction) term on the right hand side of Eq. (2.6) is proportional to the Riemann curvature scalar at radius $R$ from the Earth's centre.

To compare Eq. (2.6) with the quantum transit time, I solve the Schrödinger equation (2.2) using first order stationary perturbation theory. The resulting Lippman-Schwinger equation becomes, for small gravity gradient,

$$u(x) \approx \sin kx + \hbar^{-1}\int G(x,x')V(x')\sin kx'\,dx' \tag{2.7}$$

where

$$\psi(x,t) = u(x)e^{iEt/\hbar}. \tag{2.8}$$

(For the purposes of computing the phase change we may omit the overall normalization factor.) The stationary one-dimensional Green function in the region $x < 0$ satisfying the reflecting (vanishing) boundary condition at $x = 0$ is

$$G(x,x') = -(im/\hbar k)[e^{ik|x-x'|} - e^{-ik(x+x')}]. \tag{2.9}$$

Substituting $G$ into Eq. (2.7), and evaluating the right hand side at $x = -b$ gives

$$\begin{aligned}u(x) &\approx -\sin kb + (2m/\hbar^2 k)e^{ikb}\int_{-b}^{0}V(x')\sin^2 kx'\,dx' \\ &\approx -(1/2i)e^{ikb+i\theta} + (1/2i)e^{-ikb}\end{aligned} \tag{2.10}$$

where for small $V$ we may approximate

$$\theta \approx -(4m/\hbar^2 k)\int_{-b}^{0}V(x')\sin^2 kx'\,dx'. \tag{2.11}$$

The phase change $\varphi$ suffered by the wave function in propagating from $x = -b$ to $x = 0$ and back again to $x = -b$ is $2kb + \theta$. The $b$-dependant terms in Eq. (2.10) represent the unperturbed propagation: substitution into Eq. (1.1) yields a transit time of $2b/v$, which is identical to the classical transit time for a particle moving with uniform speed $|v| \equiv \hbar|k|$. The $\theta$-dependant term contains the additional phase change induced by the gravity gradient perturbation $V(x)$. To evaluate this term, first suppose that the particle's speed is not too low, so that $|k|b \gg 1$. Then the integral in Eq. (2.11) may be approximated:

$$\int_{-b}^{0}V(x')\sin^2 kx'\,dx' = \tfrac{1}{2}\int_{-b}^{0}V(x')\,dx' - \tfrac{1}{2}\int_{-b}^{0}V(x')\cos 2kx'\,dx' \approx \tfrac{1}{2}\int_{-b}^{0}V(x')\,dx'. \tag{2.12}$$

Inserting this approximation into Eq. (2.11) gives an additional phase shift of

$$\theta \approx -(2m/\hbar^2 k)\int_{-b}^{0}V(x')\,dx'. \tag{2.13}$$

Using Eq. (1.1), I obtain for the corresponding total quantum transit time

$$T_{\text{quantum}} \approx 2b/v + (1/Ev)\int_{-b}^{0}V(x')\,dx' \tag{2.14}$$

which is identical to the classical result Eq. (2.5), and is therefore *independent* of $\hbar$, in spite of the fact that this is a quantum mechanical result. Using the explicit form of the gravity gradient potential, Eqs. (2.3) and (2.4), one finds from Eqs. (2.13) and (2.14)

$$\theta \approx 2GMm^2b^3/3R^3\hbar^2k = \tfrac{1}{3}(GMm/R^3\hbar)b^2T_0 \tag{2.15}$$

$$T_{\text{quantum}} \approx T_0 - T_0b^2GMm/6R^3E, \tag{2.16}$$

$$\approx T_0 - T_0b^2GM/3R^3v^2. \tag{2.17}$$

The transit time Eq. (2.16) is identical to the classical transit time given by Eq. (2.6). Note that the gravity gradient phase correction $\theta$ in Eq. (2.15) depends on $\hbar$ and $m$, although the transit time given by Eq. (2.17) is independent of both $\hbar$ and $m$, consistent with the principle of equivalence.

If the particle is initially directed towards the Earth, i.e. in the $-x$ direction, and the reflecting wall placed below the 'start line,' then $T_{\text{quantum}}$, which is quadratic in the velocity $v$, is still given by Eq. (2.17). However, the phase change, Eq. (2.15), reverses in sign ($k \rightarrow -k$). The difference in phase between upward- and downward-directed particles is thus

$$\Delta\theta = \tfrac{2}{3}(GMm/R^3\hbar)b^2T_0. \tag{2.18}$$

A quantum correction arises when we take into account the neglected term

$$-\tfrac{1}{2}\int_{-b}^{0}V(x')\cos 2kx'\,dx' \tag{2.19}$$

in Eq. (2.12). Physically, this term becomes important in the slow motion limit, where the de Broglie wavelength of the particle is comparable to the distance $b$ over which the particle travels. In other words, the wavelength is long enough to 'feel' the spacetime curvature. Using the gravity gradient expression $\alpha(x+b)^2$ for $V(x)$, one finds in place of Eq. (2.17) the exact result

$$T_{\text{quantum}} = T_{\text{classical}} + T' \tag{2.20}$$

where

$$T' = -(GMb\hbar^2/R^3m^2v^5)(3 + \cos 2kb) + (2GM\hbar^3/R^3m^3v^6)\sin 2kb. \tag{2.21}$$

Two features of Eq. (2.21) should be noted. First, the additional terms depend explicitly on $\hbar$, and so represent quantum corrections to the classical transit time. Second, these terms also depend on $m$, the mass of the test particle, and so represent quantum departures from the principle of equivalence. This phenomenon was also found in Davies (2004) for the case of the particle projected vertically in the uniform gravitational potential $mgx$, if the transit time was measured from a location near the classical turning point. The result may be understood as an expression of the non-local aspect of quantum mechanics coming into conflict with the local nature of the principle of equivalence. Given this conflict, it is perhaps remarkable that, far from the turning point (or, in this paper, when the particle velocity is not too low), the quantum transit time coincides with the classical transit time, even for highly non-classical, non-local states, such as energy eigenstates. This conformity points to a deep consistency between quantum mechanics and general relativity, notwithstanding the fact that their respective conceptual bases are radically different.

To obtain some idea of the relative importance of the quantum correction, note that

$$T'/T_{\text{classical}} \sim 1/k^2 b^2, \quad (2.22)$$

so that these terms will become important when $k \sim 1/b$, i.e. when the de Broglie wavelength is $\sim b$.

## 3. Comparison with the results of Chiao and Speliotopoulos

In the foregoing I have described a radial free-fall experiment, enabling the analysis to be carried out in one space dimension. This necessitated the device of the reflecting wall to return the quantum particle to the position of the 'start line' controlling the operation of the quantum clock. It is instructive to compare my results with the related work of Chiao and Speliotopoulos (2003), who considered the more practicable scenario shown in Fig. 3. Here, a freely-falling observer on a circular orbit of the Earth observes quantum phase differences arising from the differential orbital motions along paths $A$ and $B$. Path $A$ corresponds to the upward-directed 'out-and-back' motion in my one-dimensional model, and path $B$ to the downward-directed motion, but in this case there is no need for the reflecting wall because the paths intersect naturally as a result of their orbital motion. The authors find the following expression for the phase difference between the two paths

$$\Delta\theta = (GMm/\hbar R^3) A T_0 \quad (3.1)$$

where $A$ is the shaded area shown in Fig. 3 and $T_0$ is the observer's classical transit time. Equation (3.1) is identical to my Eq. (2.18) if we make the natural identification of $A$ with $\tfrac{2}{3}b^2$.

Chiao and Speliotopoulos suggest using atom interferometry to bring the measurement of this phase difference (and, by implication, the measurement of the Peres clock transit time according to Eq. (2.16)) within the scope of experimental feasibility. The experiment should be possible with a modest improvement in sensitivity. Measurement of $\theta$ would provide a quantum mechanical means to determine spacetime curvature, and to verify that the quantum transit time (using the Peres clock definition employed in this paper) for a particle in free fall in a gravitational field coincides with the classical transit time, and thus complies with the principle of equivalence, i.e. is independent of both $\hbar$ and $m$, even though the phase change itself depends explicitly on both these parameters. However, measurement of the phase shift given by Eq. (3.1) would not demonstrate explicit quantum mechanical effects arising from the motion of the particle in the gravitational field. To fully probe the interface of quantum mechanics and gravitation would entail measurement of the quantum correction term, Eq. (2.21). This would require ultra-cold atoms and long coherence times to fulfill the condition $kb \sim 1$. Using the target values discussed by Chiao and Speliotopoulos (2003) for their proposed experiment, it seems that this condition is potentially within reach.

**References**

Amino C G, Steane A M, Bouyer P, Desbiolles, P, Dalibard J and Cohen-Tannoudji C 1993 *Phys. Rev. Lett.* **71**, 3083
Berry M V and Balazs N L 1979 *J. Phys. A: Math. Gen.* **12** 625
Chiao R Y and Speliotopoulos A D 2003 *Int. J. Mod. Phys. D* **12** 1627
Colella R, Overhauser A W and Werner S 1975 *Phys. Rev. Lett.* **34** 1472
Davies P C W 2004 *Class. Quantum Grav.* **21** 2761
Mashhoon B 1988 *Phys. Rev. Lett.* **61** 2639
van der Zouw G, Weber M, Felber J, Gähler R, Geltenbort P and A. Zeilinger A 2000 *Nucl. Instr. and Meth.* **A440** 568
Peters A, Chung K Y and Chu S 1999 *Nature* **400**, 849
Peres A 1980 *Am. J. Phys.* **48**, 552
Salecker H and Wigner E P 1958 *Phys. Rev.* **109**, 571

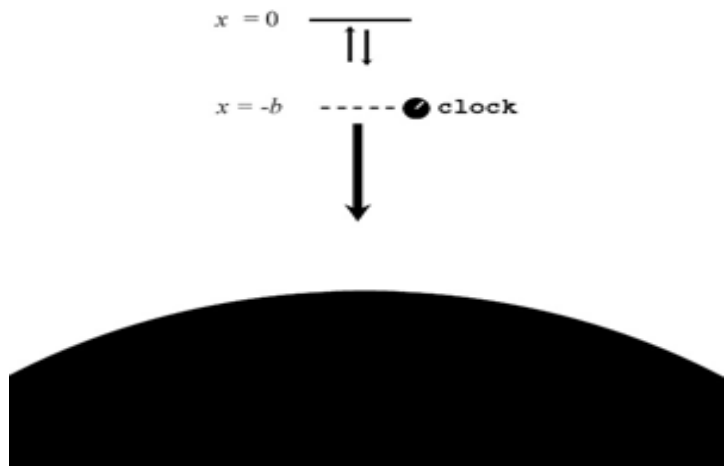

Figure 1. A quantum particle and a quantum clock are in radial free fall above the Earth. The particle is directed upwards from $x = -b$ towards a reflecting wall at $x = 0$, and its time of flight to the wall and back is measured by the quantum clock. Tidal gravitational forces will accelerate the particle relative to the clock, producing an additional phase shift in the particle's wave function.

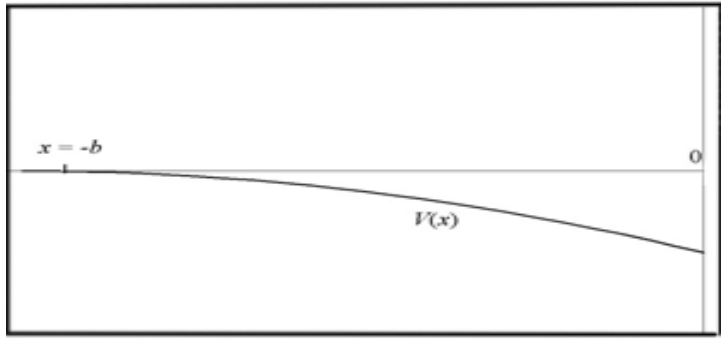

Figure 2. The tidal gravitational force acting on the particle in free fall is represented by the quadratic potential $V(x)$.

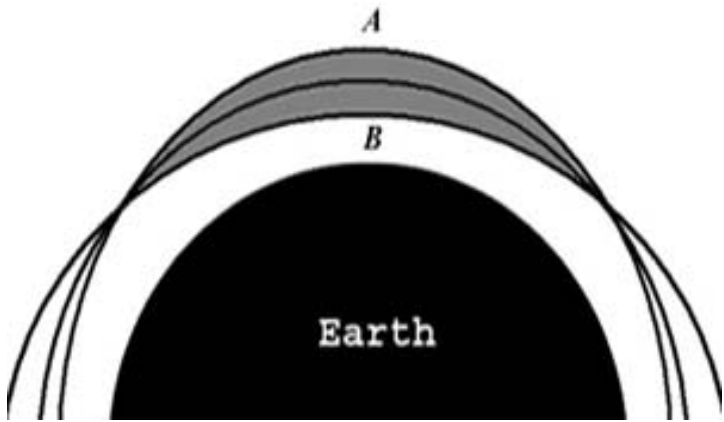

Figure 3. The Earth orbiting experiment proposed by Chiao and Speliotopoulos. Quantum particles on geodesics 1 and 2 can be used as an interferometer to measure the spacetime curvature.